\documentclass[aps,twocolumn,prl,superscriptaddress,showpacs]{revtex4}
\usepackage{epsfig, amssymb}

\begin{document}
\advance\hoffset by  -4mm

\topmargin -2 mm
\newcommand{\de}{\Delta E}
\newcommand{\mbc}{M_{\rm bc}}
\newcommand{\bb}{B{\bar B}}
\newcommand{\qq}{q{\bar q}}
\newcommand{\br}{{\cal B}}

\newcommand{\dkpi}{\bar{D}^0\to\kpi}
\newcommand{\pipi}{\pi^+\pi^-}
\newcommand{\kpi}{K^-\pi^+}
\newcommand{\kpipi}{\kpi\pi^+}
\newcommand{\trpi}{\pi^-\pi^-\pi^+}
\newcommand{\dpipi}{D\pi\pi}
\newcommand{\dspipi}{D^*\pi\pi}
\newcommand{\ddspipi}{D^{(*)}\pi\pi}

\newcommand{\DECAYDONEPDPIPI}{D_1^+ \to D^+\pi^-\pi^+}
\newcommand{\DECAYDONEZDPIPI}{{D_1^0\to D^0\pi^-\pi^+}}

\newcommand{\DECAYDONESTARPDPIPI}{D_1^+\to D^{*+}\pi^-\pi^+}
\newcommand{\DECAYDONESTARZDPIPI}{{D_1^0\to D^{*0}\pi^-\pi^+}}

\newcommand{\BDSSPI}{{B^- \to D_2^{*0}\pi^-}}
\newcommand{\BDONEPI}{{B^- \to D_1^0 \pi^-}}

\newcommand{\DECAYBNBDONEPI}{{B^-\to D_1^0 \pi^-}}
\newcommand{\DECAYBZBDONEPI}{{\bar{B}^0\to D_1^+ \pi^-}}

\newcommand{\mdpipi}{M_{D\pi\pi}}
\newcommand{\mdspipi}{M_{D^*\pi\pi}}
\newcommand{\mddspipi}{M_{D^{(*)}\pi\pi}}
\newcommand{\mpipi}{M_{\pi^+\pi^-}}
\newcommand{\mdpi}{M_{D\pi}}

\newcommand{\BNRATIOTEN}{{(1.85\pm0.29\pm0.35^{+0.0}_{-0.46})}}
\newcommand{\BZRATIOTEN}{{(0.89\pm0.15\pm0.17^{+0.0}_{-0.26})}}

\newcommand{\cpipim}{\cos \Theta(\pi^-_B\pi^-_{D^{**}})}
\newcommand{\cpipip}{\cos \Theta(\pi^-_B\pi^+_{D^{**}})}
\newcommand{\cpid}{\cos \Theta(\pi^-_B D)}
\newcommand{\done}{D_1(2420)}
\newcommand{\dtwo}{D_2^*(2460)}

\preprint{\vbox{ \hbox{BELLE-CONF-0464}
                 \hbox{ICHEP04 10-0715}
}}

\title{\Large \rm Observation of the $\done \to D\pipi$  decays. }

\affiliation{Budker Institute of Nuclear Physics, Novosibirsk}
\affiliation{Chiba University, Chiba}
\affiliation{Chonnam National University, Kwangju}
\affiliation{University of Cincinnati, Cincinnati, Ohio 45221}
\affiliation{Gyeongsang National University, Chinju}
\affiliation{University of Hawaii, Honolulu, Hawaii 96822}
\affiliation{High Energy Accelerator Research Organization (KEK), Tsukuba}
\affiliation{Hiroshima Institute of Technology, Hiroshima}
\affiliation{Institute of High Energy Physics, Chinese Academy of Sciences, Beijing}
\affiliation{Institute of High Energy Physics, Vienna}
\affiliation{Institute for Theoretical and Experimental Physics, Moscow}
\affiliation{J. Stefan Institute, Ljubljana}
\affiliation{Kanagawa University, Yokohama}
\affiliation{Korea University, Seoul}
\affiliation{Kyungpook National University, Taegu}
\affiliation{Swiss Federal Institute of Technology of Lausanne, EPFL, Lausanne}
\affiliation{University of Ljubljana, Ljubljana}
\affiliation{University of Maribor, Maribor}
\affiliation{University of Melbourne, Victoria}
\affiliation{Nagoya University, Nagoya}
\affiliation{Nara Women's University, Nara}
\affiliation{National Central University, Chung-li}
\affiliation{National United University, Miao Li}
\affiliation{Department of Physics, National Taiwan University, Taipei}
\affiliation{H. Niewodniczanski Institute of Nuclear Physics, Krakow}
\affiliation{Nihon Dental College, Niigata}
\affiliation{Niigata University, Niigata}
\affiliation{Osaka City University, Osaka}
\affiliation{Osaka University, Osaka}
\affiliation{Panjab University, Chandigarh}
\affiliation{Peking University, Beijing}
\affiliation{Princeton University, Princeton, New Jersey 08545}
\affiliation{Saga University, Saga}
\affiliation{University of Science and Technology of China, Hefei}
\affiliation{Seoul National University, Seoul}
\affiliation{Sungkyunkwan University, Suwon}
\affiliation{University of Sydney, Sydney NSW}
\affiliation{Tata Institute of Fundamental Research, Bombay}
\affiliation{Toho University, Funabashi}
\affiliation{Tohoku Gakuin University, Tagajo}
\affiliation{Tohoku University, Sendai}
\affiliation{Department of Physics, University of Tokyo, Tokyo}
\affiliation{Tokyo Metropolitan University, Tokyo}
\affiliation{Tokyo University of Agriculture and Technology, Tokyo}
\affiliation{University of Tsukuba, Tsukuba}
\affiliation{Virginia Polytechnic Institute and State University, Blacksburg, Virginia 24061}
\affiliation{Yonsei University, Seoul}
  \author{K.~Abe}\affiliation{High Energy Accelerator Research Organization (KEK), Tsukuba} 
  \author{K.~Abe}\affiliation{Tohoku Gakuin University, Tagajo} 
  \author{I.~Adachi}\affiliation{High Energy Accelerator Research Organization (KEK), Tsukuba} 
  \author{H.~Aihara}\affiliation{Department of Physics, University of Tokyo, Tokyo} 
  \author{M.~Akatsu}\affiliation{Nagoya University, Nagoya} 
 \author{D.~Anipko}\affiliation{Budker Institute of Nuclear Physics, Novosibirsk} 
  \author{Y.~Asano}\affiliation{University of Tsukuba, Tsukuba} 
  \author{T.~Aushev}\affiliation{Institute for Theoretical and Experimental Physics, Moscow} 
  \author{T.~Aziz}\affiliation{Tata Institute of Fundamental Research, Bombay} 
  \author{S.~Bahinipati}\affiliation{University of Cincinnati, Cincinnati, Ohio 45221} 
% ...
\collaboration{The Belle Collaboration}

\begin{abstract}
We report on the first observation of $\done\to D\pipi$
decays (where the contribution from the dominant known $D_1 \to D^* \pi$ decay mode
is excluded) in the $B\to D_1\pi$ decays. The observation is based on $15.2\times 10^7 \bb$ events
collected with the Belle detector at the KEKB collider. We also set
90\% confidence level upper limits for the
$D_2^*\to D^{(*)}\pipi$ and
$D_1 \to D^*\pipi$ decays in the $B\to \dtwo\pi$ and $B\to \done\pi$ decays, respectively.
\end{abstract}
\pacs{13.25.Hw, 14.40.Lb}
\maketitle

\tighten

{\renewcommand{\thefootnote}{\fnsymbol{footnote}}}
\setcounter{footnote}{0}


The $D_1 \to D^* \pi$ decay  is currently known to be the primary decay
mode of the $\done$ meson~\cite{PDG}. However, the transitions $D_1\to D\pi\pi$ via other
intermediate quasi-two-body resonance states or via non-resonant decays 
are possible and may contribute to the $D_1$ total width. Measurements
of their branching ratios and analysis of the decay dynamics are
particularly relevant to a study of production rates of various
$D^{**}$ excitations in $B$ decays.

The ratio of the branching fractions
$R={\cal B}(B^-\to D^{*0}_2\pi^-)/{\cal B}(B^-\to D^0_1\pi^-)$
is calculated in HQET and the factorization approximation in
Refs.~\cite{leg,neubert}. In Ref.~\cite{leg}, $R$ is found to depend
on the values of the subleading Isgur-Wise functions ($\hat{\tau}_{1,2}$)
describing $\Lambda_{QCD}/m_c$ corrections; thus measurement of $R$ can be 
used to estimate the sub-leading functions.
In  Ref.~\cite{neubert}, some of the subleading terms
are estimated and the ratio is determined to be  
$R\approx 0.35\biggl|(1+\delta_8^{D2})/(1+\delta_8^{D1})\biggr|^2$,
where $\delta_8^{D1(D2)}$ are non-factorizable corrections that are
expected to be small.

The first observation of $D^{**}$ (denotes $P$-wave $D$ excitations) production in $B$ decays was
reported by CLEO~\cite{CLEO9625}. From their studies and the
measurement of the ratio
$\br(D^{*0}_2\to D^{+}\pi^-)/\br(D^{*0}_2\to D^{*+}\pi^-)$~\cite{CL2,AR3}
the $R$ value was determined to be $R=1.8\pm0.8$, where it was assumed
that decays of $D_1$ and $D_2^*$ mesons are saturated by the two-body
$D\pi,~D^*\pi$ modes. Recently the branching fractions for the decays
$B\to D^{**}\pi \to D^{(*)}\pi\pi$ have been measured with better
accuracy~\cite{kuzmin}, resulting in $R=0.77\pm0.15$. The existence of
$D_{(1,2)}$ decay channels other than $D_1\to D^{(*)}\pi$ can also
affect the $R$ value, either decreasing or increasing the currently
observed $2.8\sigma$ difference between the prediction and
experimental results.

In this Letter we report the first observation of the  $\DECAYDONEPDPIPI$ and $\DECAYDONEZDPIPI$ decays.
The $D_1$ mesons were reconstructed from the $\DECAYBZBDONEPI$ and $\DECAYBNBDONEPI$ decays, respectively.
The results are based on a sample of $15.2\times 10^7$
$\bb$ pairs produced at the KEKB asymmetric energy $e^+e^-$
collider~\cite{KEKB}. The inclusion of charge conjugate states
is implicit throughout this report.

The Belle detector has been described elsewhere~\cite{NIM}.
Charged tracks are selected with a set of requirements based on the
average number of hits in the central drift chamber (CDC) and on the
distance of the closest approach to the interaction point. Track
momentum transverse to the beam axis of at least 0.05~GeV$/c$ is
required for all tracks in order to reduce the combinatorial background. 
For charged particle identification (PID), the combined information
from specific ionization in the CDC ($dE/dx$),
time-of-flight scintillation counters and aerogel \v{C}erenkov
counters is used. Charged kaons are selected with PID criteria that
have an efficiency of 88\%, a pion misidentification probability of
8\%, and negligible contamination from protons. All charged tracks
with PID responses consistent with a pion hypothesis that are not
positively identified as electrons are considered as pion candidates.
Photon candidates are selected from calorimeter showers not associated
with charged tracks. An energy deposition of at least 30~MeV and a
photon-like shape are required for each candidate. Pairs of photons
with an invariant mass within 12~MeV$/c^2$ ($\sim 2.5\sigma$) of the
$\pi^0$ nominal mass~\cite{PDG} are considered as $\pi^0$ candidates.

We reconstruct $D^0 (D^+)$ mesons in the $\kpi$ ($\kpipi$) decay
channel and require the invariant mass to be within 15~MeV$/c^2$
($\sim 3\sigma$) of the $D^0 (D^+)$ mass. Then, $D^{*0} (D^{*+})$ mesons
are reconstructed in the $D^0\pi^0 (D^0\pi^+)$ decay mode. The
calculated mass difference between $D^{*0} (D^{*+})$ and $D^0$
candidates is required to be within 2~(1.5)~MeV$/c^2$
($\sim 2.5\sigma$) of the expected value~\cite{PDG}.
For $D^*\to D^0\pi$ decays the $D^0\to K^-\pi^+\pi^+\pi^-$ mode is
also included (the same $D^*$ parameters were used as above).

\begin{table*}[t]
\caption{Number of events, efficiencies and branching fraction products of $B \to  D^{**}\pi,
         D^{**} \to D^{(*)}\pipi$  decays.
         }
\medskip
\label{fitresults}
  \begin{tabular*}{\textwidth}{l@{\extracolsep{\fill}}|c|c|c|c}\hline
 Mode  &  $N_{sig}$ & $\varepsilon$ ($10^{-2}$) & 
 ${\cal B}$ ($10^{-4}$) & Significance\\\hline

$\DECAYBNBDONEPI$, $\DECAYDONEZDPIPI$ &  $151\pm24$ & 
14.1 & $\BNRATIOTEN$
& $8.7\sigma$ \\
$\DECAYBZBDONEPI$, $\DECAYDONEPDPIPI$ & $124\pm20$ & 
9.9 & $\BZRATIOTEN$ & $10\sigma$ \\
$B^- \to D_1^{0} \pi^-,D_1^{0}\to D^{*0}\pipi$ & $< 1.2$ & 2.2 & $< 0.06$ & -\\
$\bar{B}^0 \to D_1^{+} \pi^-,D_1^{+}\to D^{*+}\pipi$ & $< 12.0$ & 3.4 & $< 0.33$ & -\\
$B^- \to D_2^{*0} \pi^-,D_2^{*0}\to D^{*0}\pipi$ & $< 4.4$ & 2.2 & $<0.22$ & -\\
$\bar{B}^0 \to D_2^{*+}\pi^-,D_2^{*+}\to D^{*+}\pipi$ & $< 9.0$ & 3.4 & $<0.24$ & -\\
\hline

  \end{tabular*}
\end{table*}

We combine $D^{(*)}$ candidates with $\trpi$ to form $B$ mesons.
Candidate events are identified by their center-of-mass (CM) energy
difference, \mbox{$\de=(\sum_iE_i)-E_{\rm beam}$}, and the beam
constrained mass, $\mbc=\sqrt{E_{\rm beam}^2-(\sum_i\vec{p}_i)^2}$,
where $E_{\rm beam}$ is the beam energy and $\vec{p}_i$ and $E_i$
are the momenta and energies of the decay products of the $B$ meson
in the CM frame. We define the signal region as
$5.273$~GeV$/c^2<\mbc<5.285$~GeV$/c^2$ and $|\de|<25$~MeV.
The sidebands are defined as
$5.273$~GeV$/c^2<\mbc<5.285$~GeV$/c^2$ and $25$~MeV$<|\de|<50$~MeV.
If there is more than one $B$ candidate in an event, the one
with $D^{(*)}$ mass closest to the nominal value and the best
$\pi^-\pi^-\pi^+$ vertex is chosen. We use Monte Carlo (MC) simulation
to model the detector response and determine the acceptance~\cite{GEANT}.

Variables that characterize the event topology calculated in the CM frame are used to suppress 
background from the two-jet-like $e^+e^-\to\qq$ continuum process.
We require $|\cos\theta_{\rm thr}|<0.80$, where $\theta_{\rm thr}$ is 
the angle between the thrust axis of the $B$ candidate and that of the 
rest of the event; this eliminates 77\% of the continuum background
while retaining 78\% of the signal events. We also define a Fisher
discriminant, ${\cal F}$, which is based on the production angle of
the $B$ candidate, the angle of the thrust axis with respect to the
beam axis, and nine parameters that characterize the momentum flow
in the event~\cite{VCal}. We impose a requirement on ${\cal{F}}$ that
rejects 67\% of the remaining continuum background and retains 83\%
of the signal.

To suppress the large contribution from the dominant
$D_1\to D^*\pi\to D\pi\pi$ decay mode we apply a requirement on
the invariant mass of the relevant $D\pi$ combination
$|(m_{D\pi} - m_{D}) - (m^{PDG}_{D^*}-m^{PDG}_{D})|>6$
MeV/$c^2$~\cite{PDG} (10$\sigma$).

\begin{figure}[!ht]
\begin{center}

  \includegraphics[width=0.23\textwidth,height=0.15\textheight] {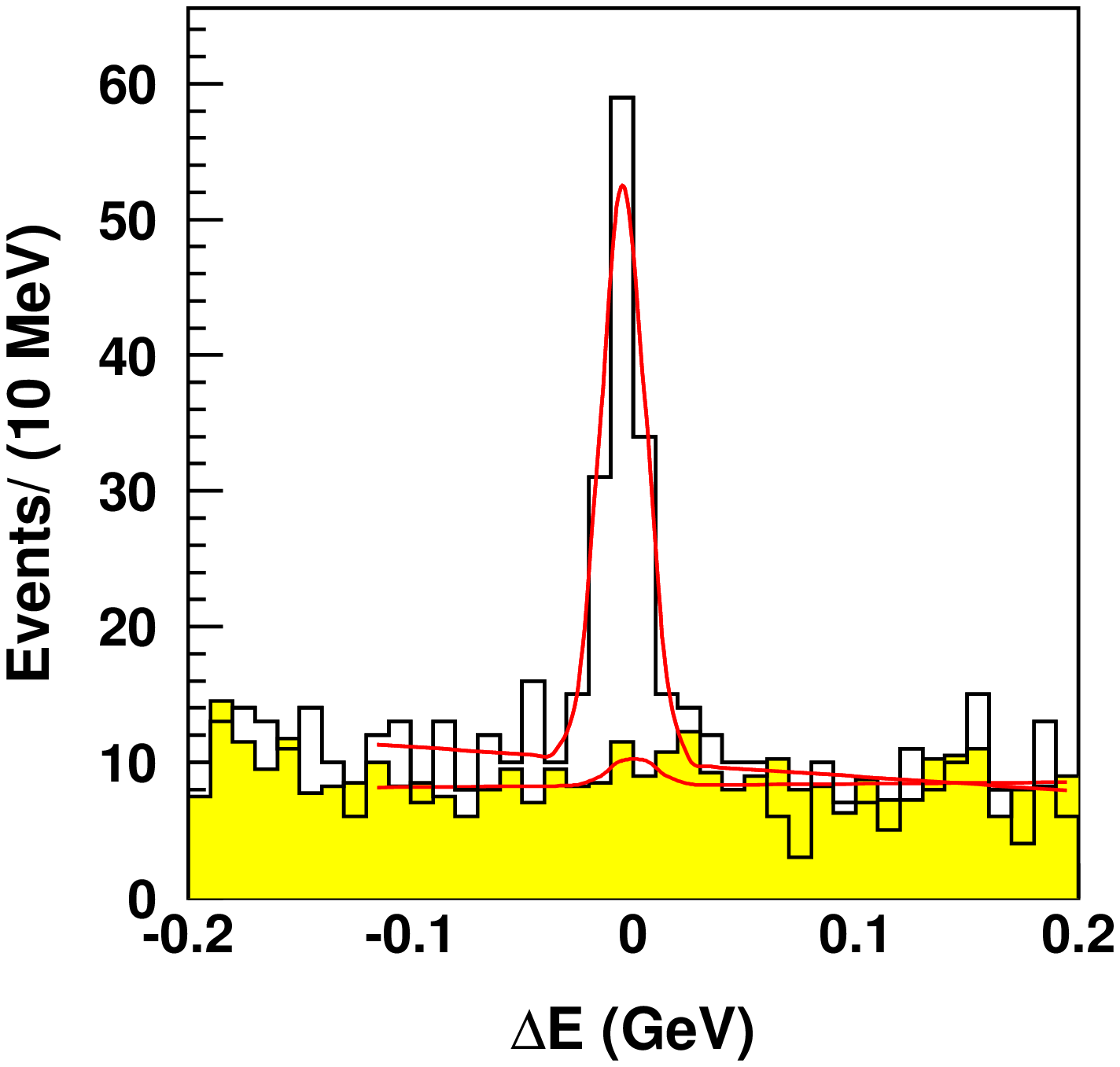} 
  \includegraphics[width=0.23\textwidth,height=0.15\textheight] {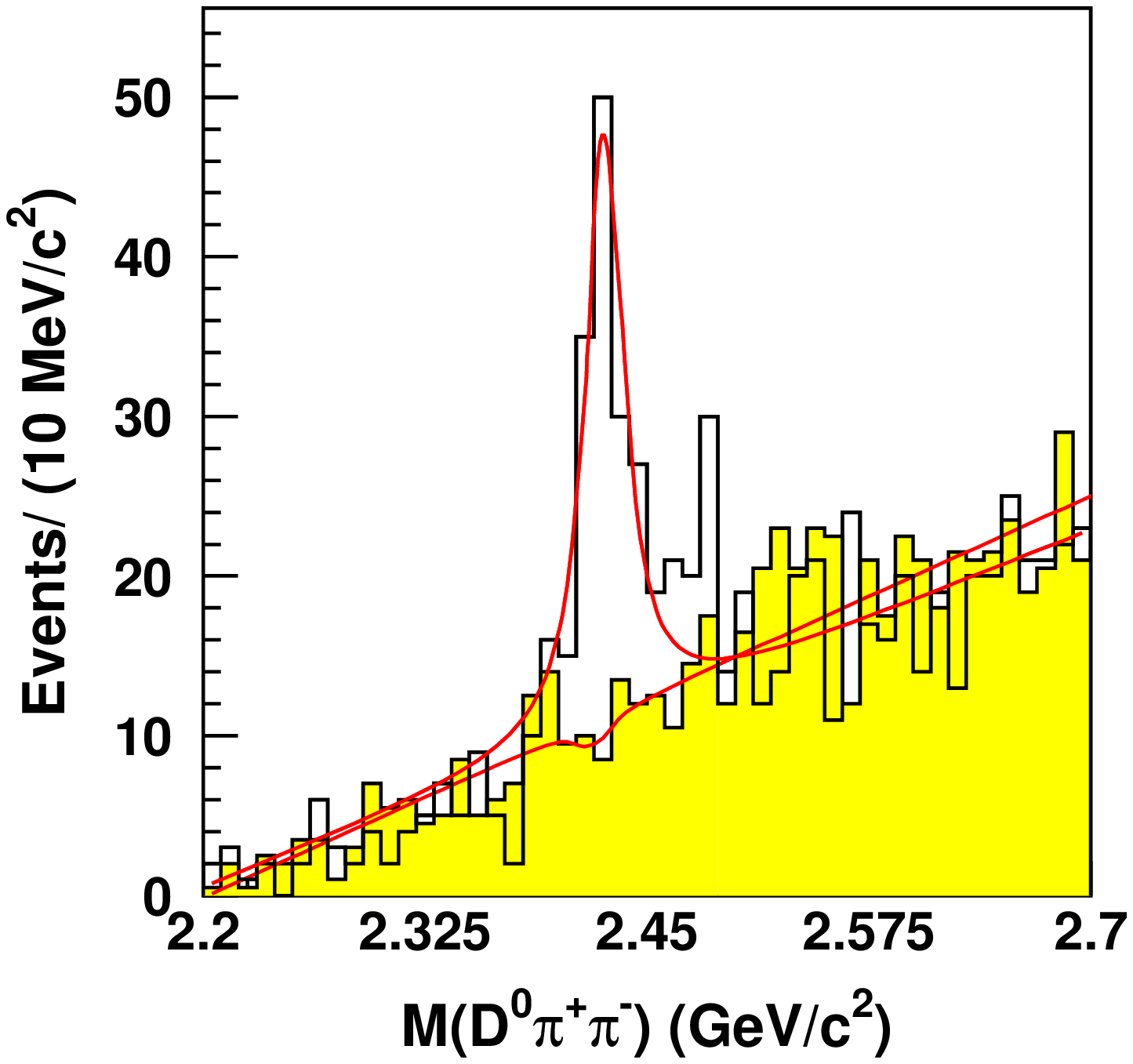}\\ 
  \includegraphics[width=0.23\textwidth,height=0.15\textheight] {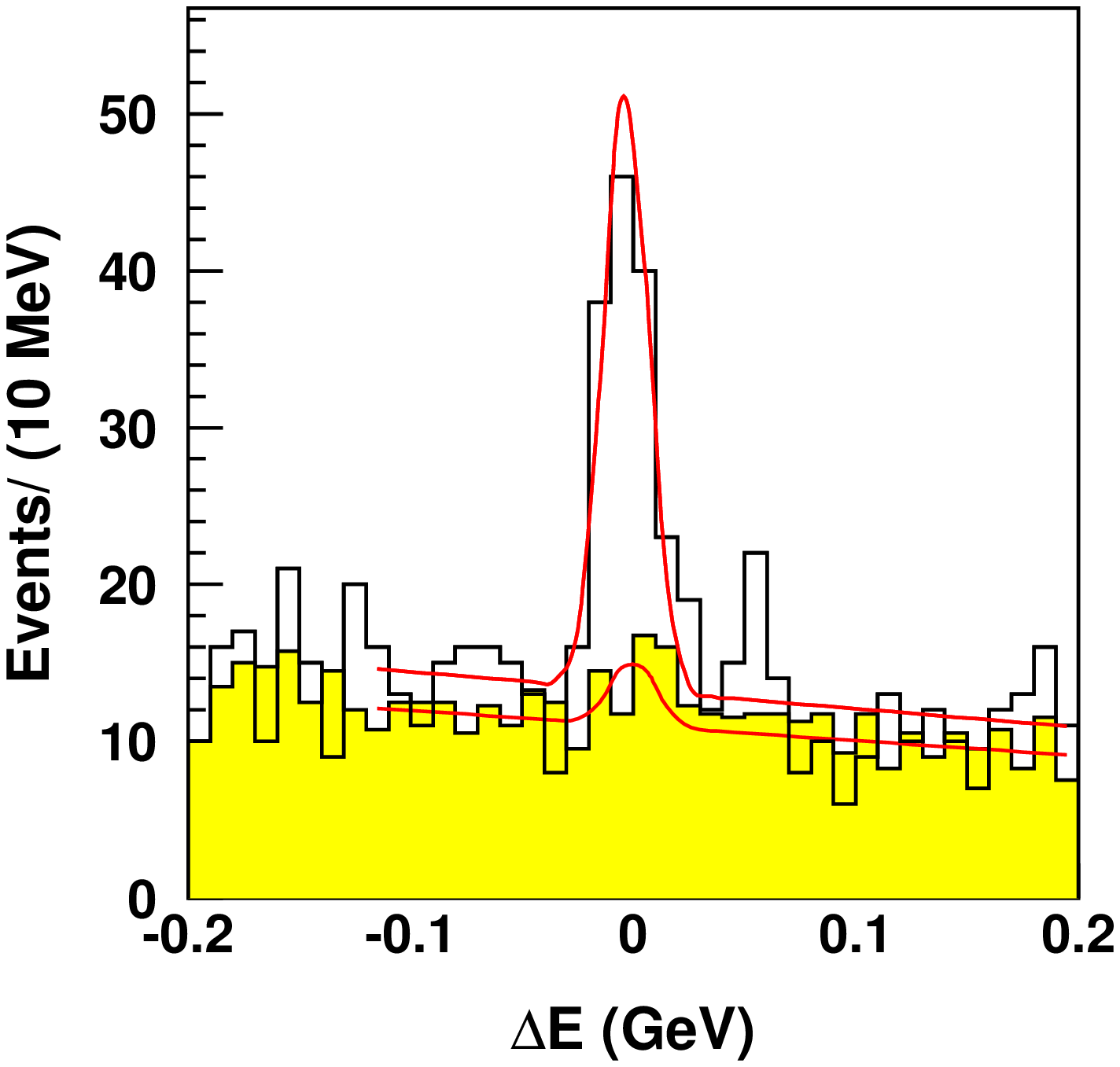} 
  \includegraphics[width=0.23\textwidth,height=0.15\textheight] {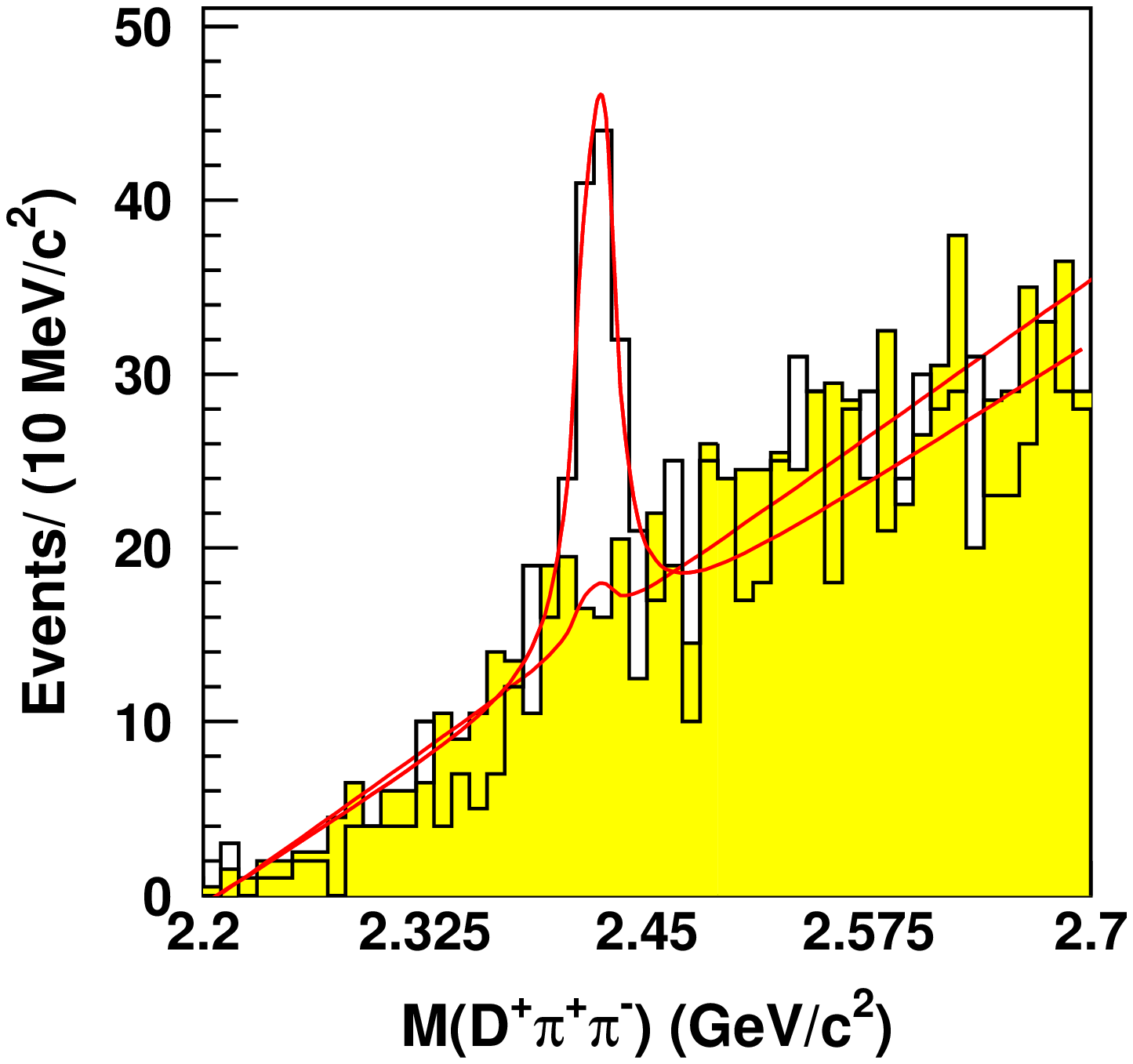}\\ 
  \includegraphics[width=0.23\textwidth,height=0.15\textheight] {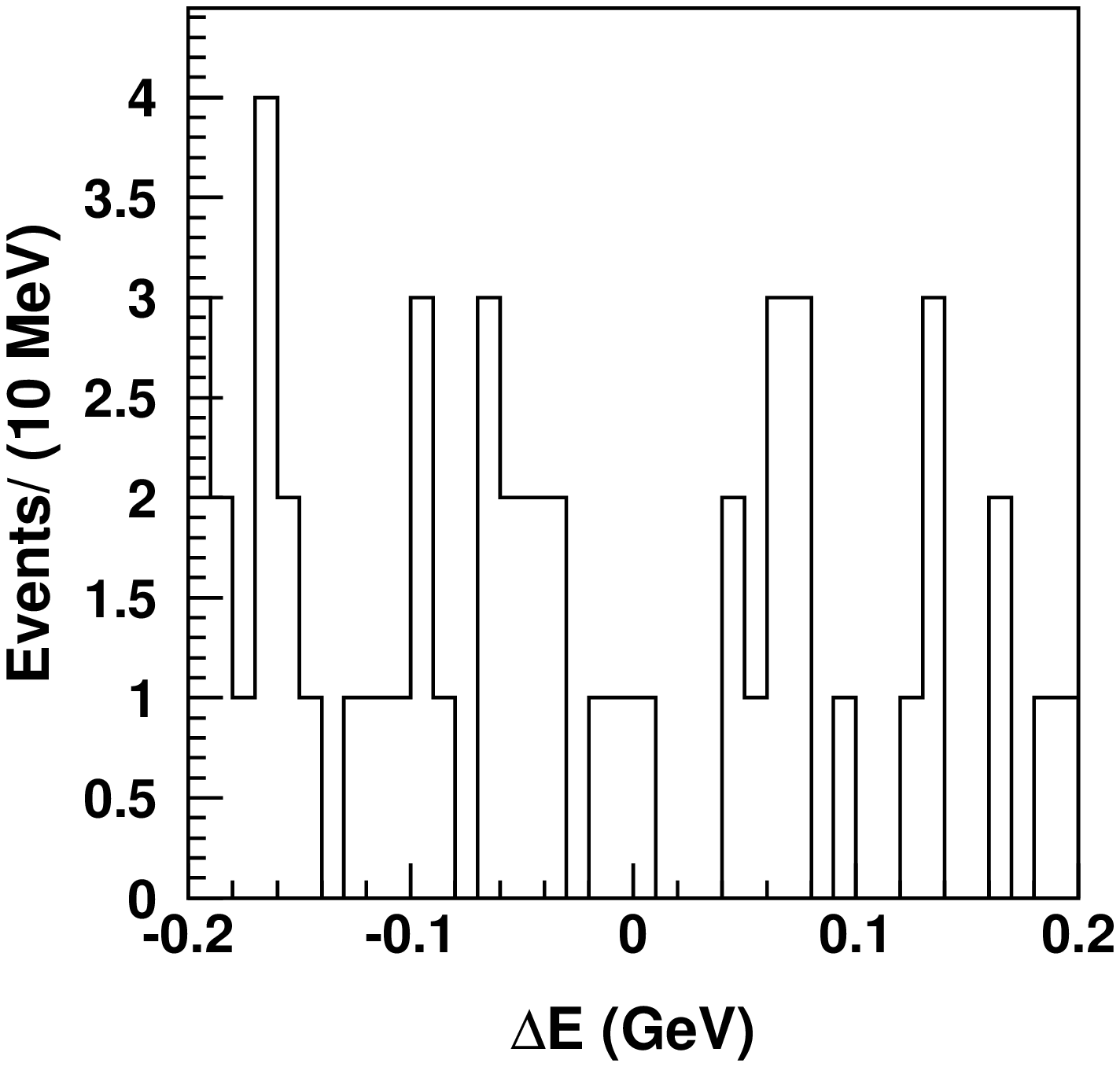} 
  \includegraphics[width=0.23\textwidth,height=0.15\textheight] {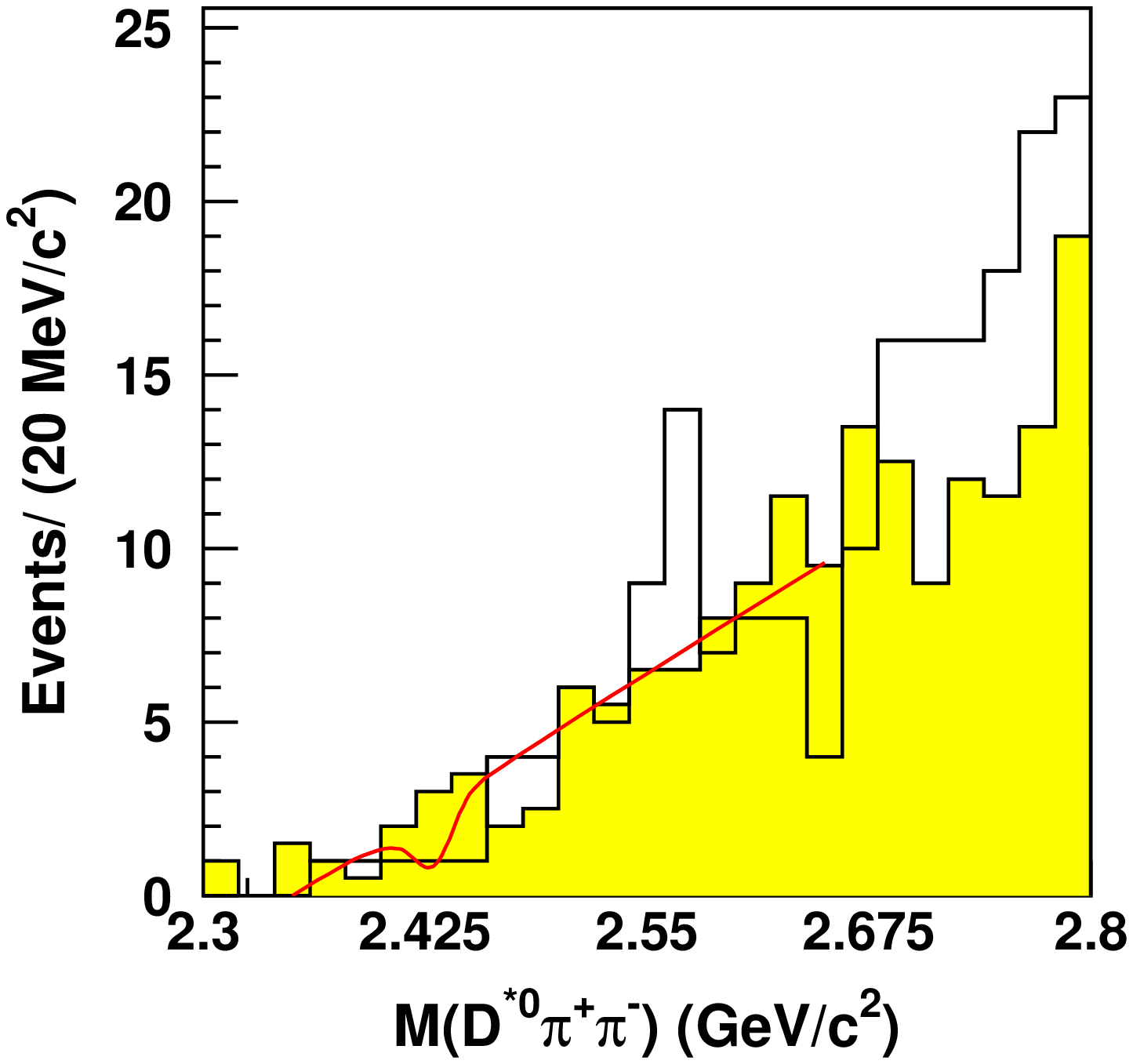}\\
  \includegraphics[width=0.23\textwidth,height=0.15\textheight] {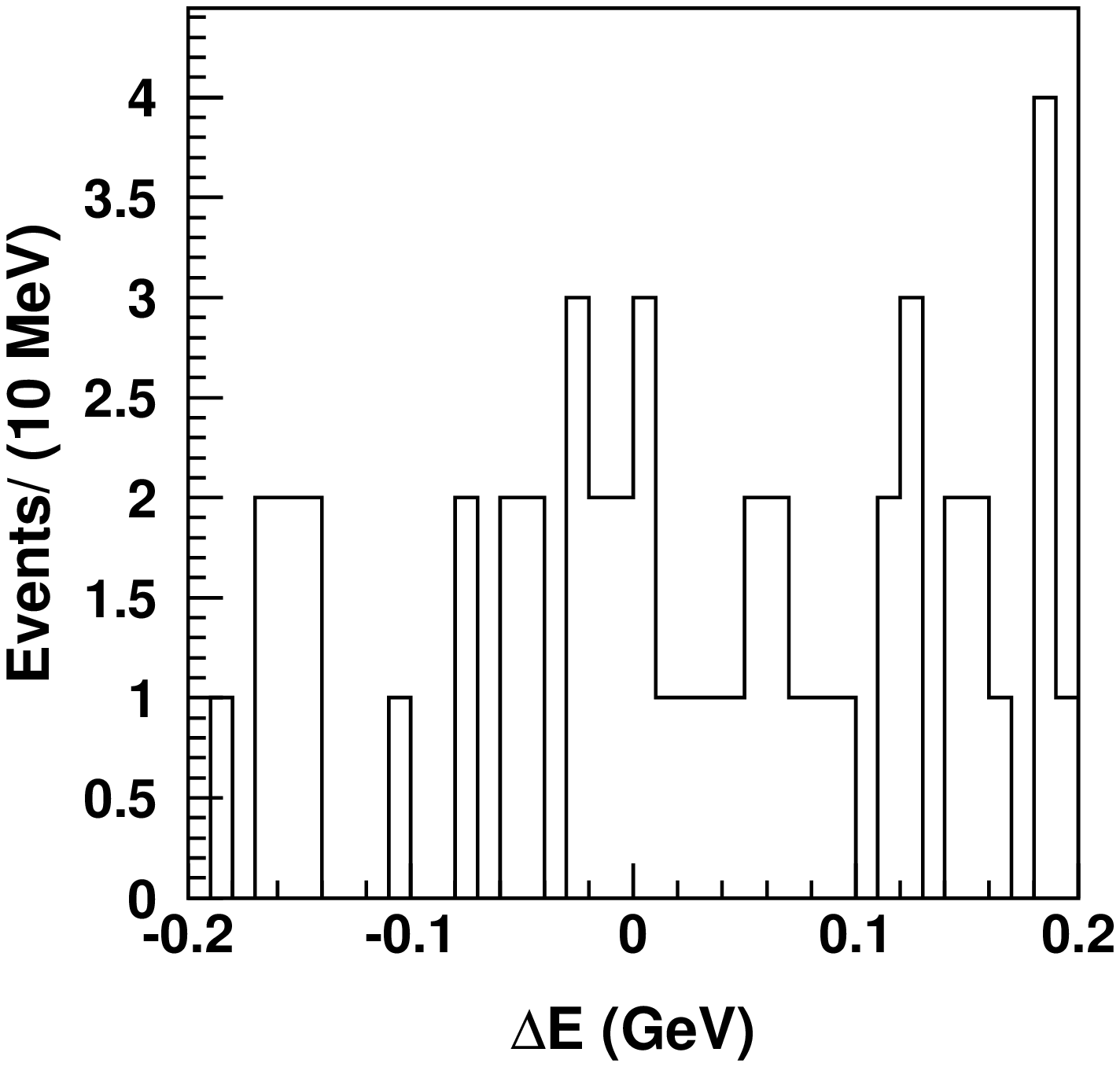} 
  \includegraphics[width=0.23\textwidth,height=0.15\textheight] {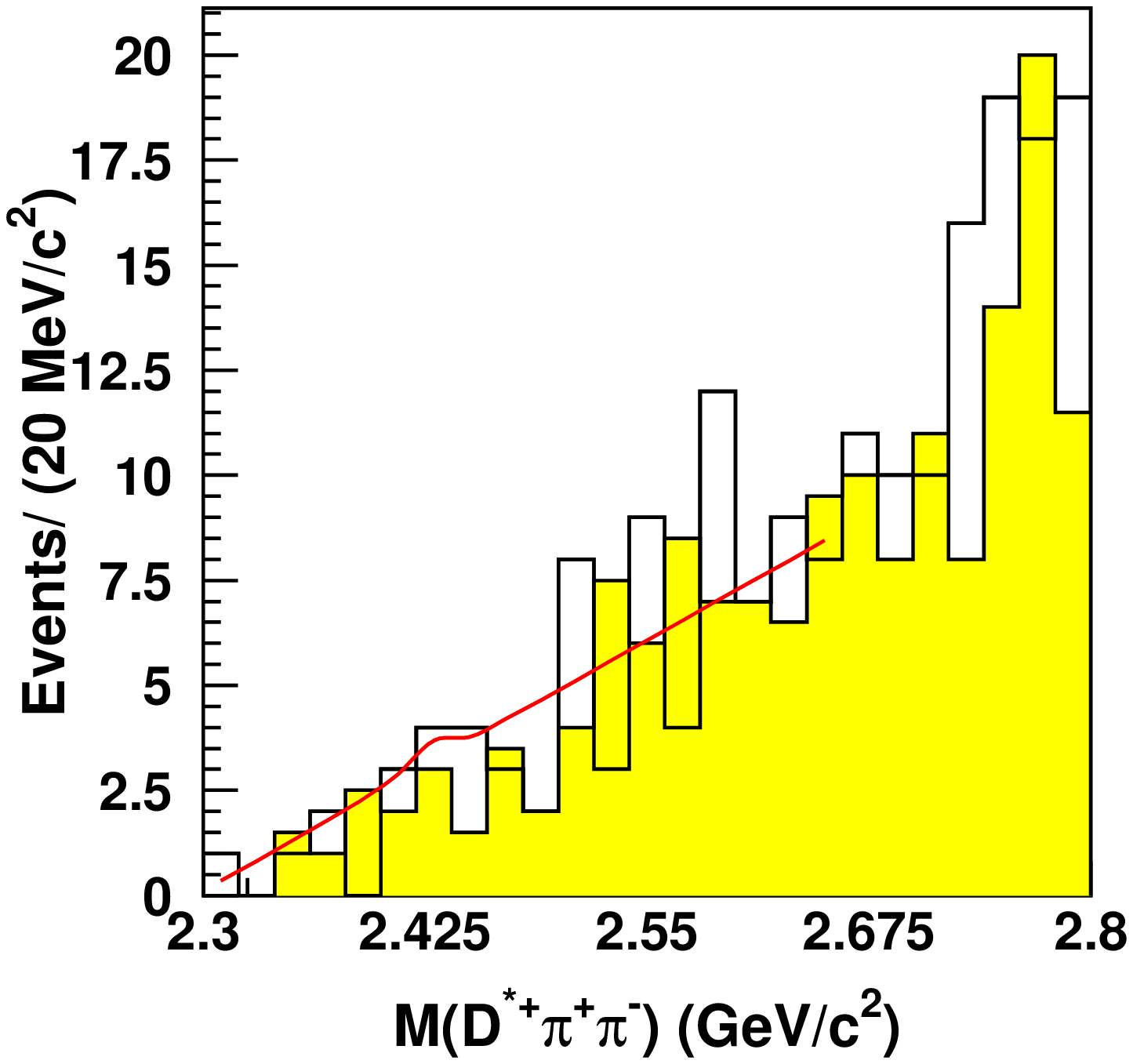}\\ 
\end{center}
  \caption{$\de$ (left) and $\mdpipi$ (right) distributions for the
  $\DECAYDONEZDPIPI$ (first row), $\DECAYDONEPDPIPI$ (second row),
$\DECAYDONESTARZDPIPI$ (third row) $\DECAYDONESTARPDPIPI$ (fourth row).
    Open histograms
    represent the data from the signal area, hatched histograms show
    the $\mdpipi$ (where applicable) and $\de$ sidebands, respectively, 
the curves are the fit results - for the signal area and sidebands.}
  \label{demdpipi}
\end{figure}

The $\de$ and $\mddspipi$ distributions for the selected
$B\to D_1\pi$, $D_1\to\ddspipi$ candidates are shown in
Fig.~\ref{demdpipi}. To plot the $\de$
distributions, we require $\mbc$ to lie in the signal region
with an additional requirement $|\mddspipi-M_{D_1}|<25$~MeV/$c^2$,
where $M_{D_1}$ is the $D_1$ world average mass value; for the
$\mddspipi$ distributions we select events from the $\de$
signal region. (Although there are two $D\pi^+\pi^-$ combinations,
they are kinematically separated in the $D_1$ mass region.)
Clear signals are observed for $\DECAYBNBDONEPI$, $\DECAYDONEZDPIPI$
and $\DECAYBZBDONEPI$, $\DECAYDONEPDPIPI$ decays. For branching
fraction calculations we use signal yields determined from the fit
to $\mdpipi$ distributions as it allows us to directly estimate a
possible contribution from $B\to D_2\pi$, $D_2\to D\pi\pi$ decay. 
The signal shape distribution is parameterized by a convolution
of a resolution Gaussian ($\sigma=2.5$ MeV/$c^2$) with a signal
Breit-Wigner function; the background is represented by a linear
function. The $D_1$ mass and width determined from the fit are 
$M_{D^0_1}=2426\pm 3\pm 1$ MeV/$c^2$ (statistical and systematic error, respectively), $\Gamma_{D^0_1}=24\pm 7\pm 8$ MeV/$c^2$
for $D_1^0$ and  $M_{D^{+}_1}=2421\pm 2\pm 1$ MeV/$c^2$,
$\Gamma_{D^+_1}=21\pm 5\pm 8$ MeV/$c^2$ for $D^+_1$; these are consistent
with the world average values~\cite{PDG}. The signal yields are
given in Table~\ref{fitresults} (in the cited branching ratios the first and second errors are statistical and systematic; where a third error given is due to model uncertainty). For the $B\to D_1\pi \to D^*\trpi$
decay channels, we do not observe statistically significantly signals and
thus determine 90\% CL upper limits~\cite{Feldman} for their
branching fractions. In the fit to the $\mdspipi$ distribution, we fix
the $D_1$ mass and width at their world average values. The
statistical significance of signals quoted in Table~\ref{fitresults}
is defined as $\sqrt{-2\ln(L_0/L_{\rm max})}$, where $L_{\rm max}$
and $L_0$ denote the maximum likelihood with the nominal signal
yield and with the signal yield fixed at zero, respectively.

To account for contamination from other possible $D_1$ production
mechanisms (such as $e^+e^-\to c\bar{c}$ continuum production or
semileptonic  $B\to D_1 l \bar{\nu}$ decays), we fit the $\mdpipi$
distribution for events in the $\de$ sidebands. In this fit,
we fix the $D_1$ mass and width at their world average values.
The fits give $-6\pm8$ events for the $D_1^0$ and $10\pm11$ events for
the $D_1^+$.

The $B\to \dtwo \pi, D_2^*\to D\pipi$ decay may also contribute 
to the $B\to D\trpi$ final state. To analyse a possible effect,
we made a simultaneous fit of the $M(D^0_1\pi^+\pi^-)$ and
$M(D^+_1\pi^+\pi^-)$ distributions, where we assume isospin
invariance and require the ratio $N({D_2^*})/N({D_1})$
to be the same for both charge combinations. The fit finds the
ratio $N(D_2^*)/N(D_1)=0.33\pm0.14$ and 
signal yields of $N(D_1^0)=120\pm17$, $N(D_1^+)=107\pm16$.
Thus, we set a 90\% CL upper limit for $D_2^*$: 
$\br(B\to D_2^{*}\pi^-)\times{\cal B}(D_2^{*}\to D\pipi)
<0.55 \br(B\to D_1\pi^-)\times{\cal B}(D_1\to D\pipi)$~\cite{footnote1}.
The number of events for the $D_1$ mechanism
obtained in the fit with the $\frac{D_2^*}{D_1}$ ratio fixed to 0.46
gives the model uncertainty for the
$D_1$ yield: $^{+0}_{-21}\%$.
Another asymmetric uncertainty, coming from the other possible $D_1$
sources,
is $^{+0}_{-10}\%$ for $D_1^{0}$ and  $^{+0}_{-22}\%$ for $D_1^{\pm}$.
These two uncertainties are combined in the final resutls as ''model'' uncertainty.

The signal yields extracted from the $\de$ distributions are used only for
a consistency check of the results. The $\de$ signal shape is
parameterized by a Gaussian with parameters determined from signal
MC. The $\de$ background shape is described by a linear function.
We restrict the fit to the range $-0.1$~GeV~$<\de<0.2$~GeV to avoid
contributions from other $B$ decays, where an additional pion
is not reconstructed. Signal yields obtained from the fits to
$\de$ distributions are $106\pm12$ for $D^0\pi^+\pi^-$ and $96\pm13$
for $D^+\pi^+\pi^-$, while the corresponding reconstruction
efficiencies are 10.8\% and 7.6\%, respectively. Thus the obtained
reconstructed event numbers by the two methods are consistent within
the statistical uncertainty. 

\begin{figure}[!ht]
 \includegraphics[width=0.23\textwidth,height=0.15\textheight] {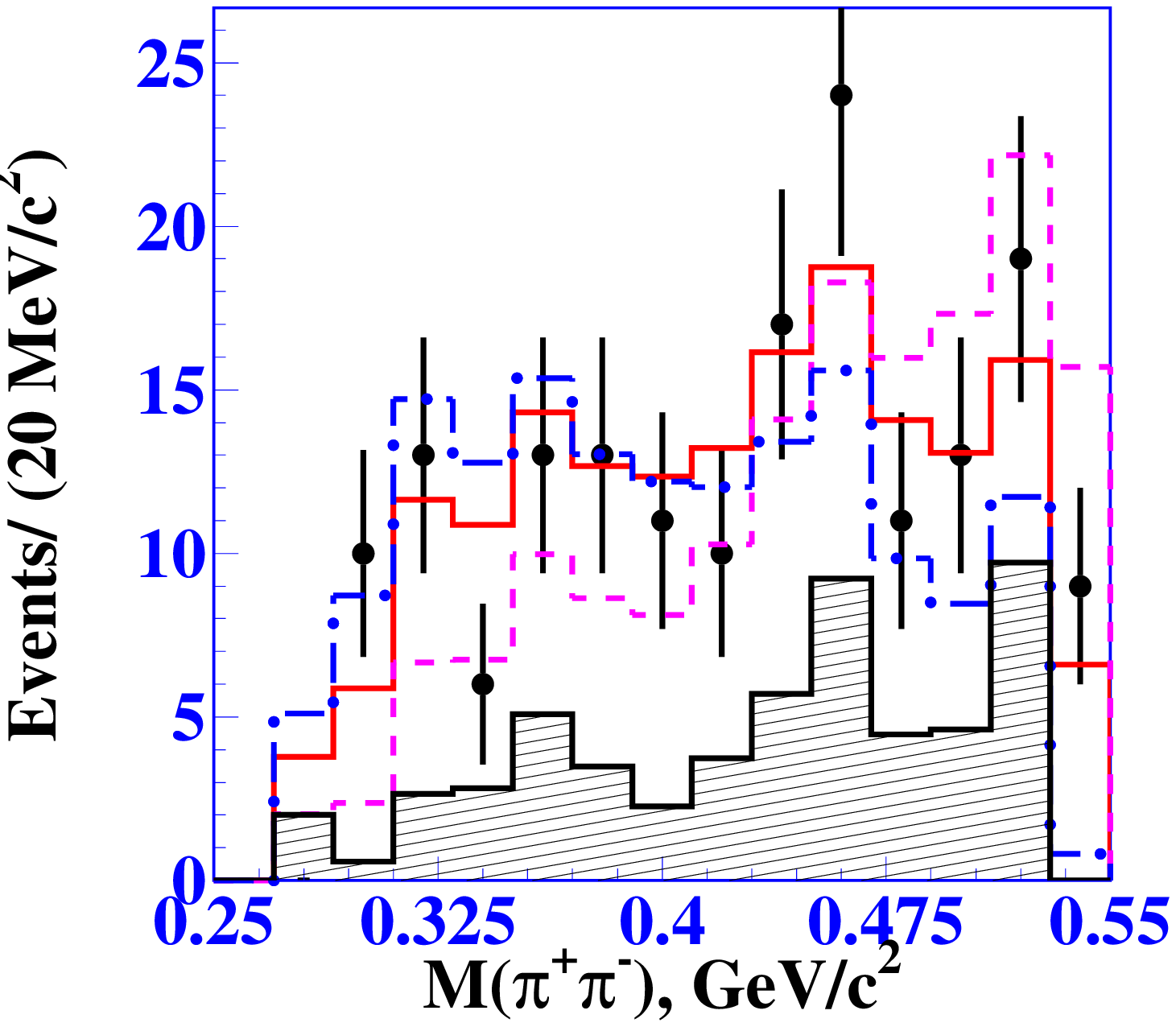} 
  \includegraphics[width=0.23\textwidth,height=0.15\textheight] {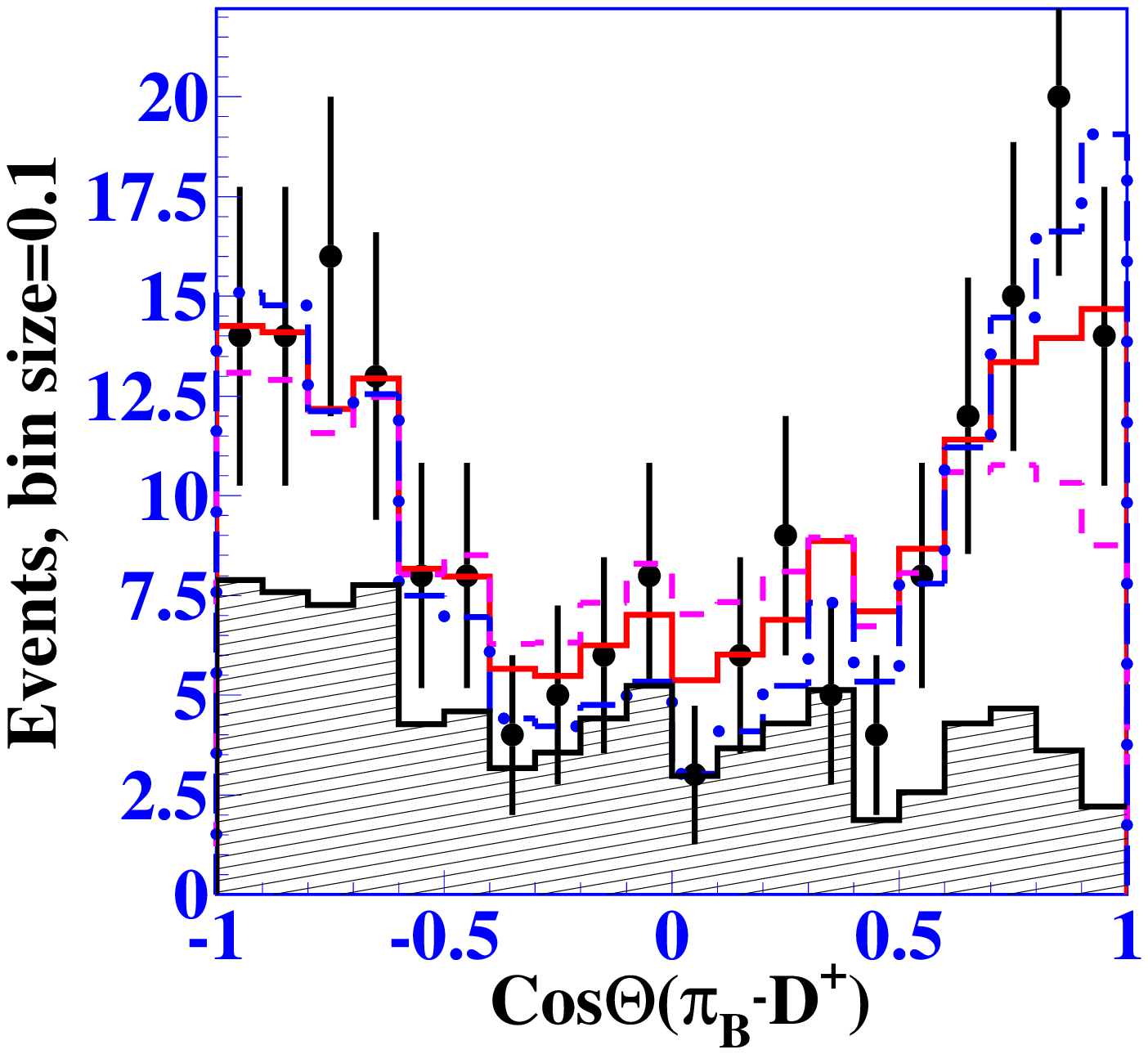} 
  \caption{$\mpipi$ (left) and $\cpid$ (right) distributions for the
           $\DECAYDONEZDPIPI$ and $\DECAYDONEPDPIPI$, respectively. Points with error bars represent the
           experimental data, solid line - $D^{*0}\pi$,
           dashed - $D\rho$, chain
           - $D f_0$ models with the expected background added. 
           The hatched histogram corresponds to expected background 
	   (from $\de$ sidebands). }
  \label{modelfit}
\end{figure}

In order to determine the total  $D_1\to D\pi\pi$ width, analysis
of final states with neutral pions is required. With only the
$D_1\to D\pi^+\pi^-$ branching fraction measurement, the analysis
of the decay dynamics could also be useful to determine the total
$D_1\to D\pi\pi$ width. As the limited statistics do not allow us to
perform the full amplitude analysis, we consider the one-dimensional
projections of several variables: $\mdpi$,
$\mpipi$, $\cpipim$, $\cpipip$, and $\cpid$ (where all angles are
calculated in the $D^{**}$ rest frame). Although these
variables are not independent, they highlight each model's
features. For instance, the helicity angle distributions differentiate
 between the $D_1\to D(\pi\pi)$ and $D_1\to (D\pi)\pi$
models. We select events from the $B$ signal region
with the additional requirement $|M(D\pi\pi)-M_{D_1}|<25$ MeV/$c^2$. 
Decays through the following quasi-two-body intermediate states
are considered:
$D_1\to D\rho^0 \to D\pi^+\pi^-$,
$D_1\to D^*_0(2308)\pi\to D\pi\pi$ and 
$D_1\to D f_0(600)\to D\pi^+\pi^-$
(we set $M_{f_0}=0.8$~GeV/$c^2$ and
$\Gamma_{f_0}=0.8$~GeV/$c^2$; the $D^*_0(2308)$ parameters are
taken from Ref.~\cite{kuzmin}). We use the simplest non-trivial
Lorentz-invariant expressions for the corresponding matrix elements
in MC simulation~\cite{footnote2}.
We fit the experimental data with different models.
For each variable we plot two distributions: one from 
the signal region and the other from the $\de$ sideband. 
We perform a simultaneous
fit to these distributions, assuming a Poisson-like profile in each bin whose
mean is the sum of the background and signal (for a given model)
in the signal region or the background only in the sideband.
The obtained differences of likelihood values 
for all variables are listed in Table~\ref{d1study_chisq}.
Figure~\ref{modelfit} shows the $\mpipi$ and $\cpid$
distributions along with expectations based on 
different $D_1\to D\pi^+\pi^-$ decay models~\cite{footnote3}.
Although the $D_1\to D^*_0 \pi$ decay mechanism describes the data best,
some contribution from other mechanisms cannot be excluded completely. 

\begin{table}[t]
\caption{Comparison of goodness-of-fit tests for the considered 
$D_1\to D\pi\pi$ decay models.}
\medskip
\label{d1study_chisq}
\begin{tabular}{l|c|c}\hline
Distribution & 
$(-2\ln{L_{D\rho}/L_{D^{*}_0\pi}})^{1/2}$ &
$(-2\ln{L_{D f_0}/L_{D^{*}_0\pi}})^{1/2}$\\\hline
 & $D^0$ , $D^+$ &  $D^0$ , $D^+$\\\hline\hline
$\mdpi$ &  $2.5$ , $2.7$ & $1.9$ , $2.9$ \\\hline
$\mpipi$  & $3.4$ , $1.6$ & $5.2$ , $4.7$\\\hline
$\cpipip$  & $1.6$ , $2.5$ & $-2.0\footnote{In this case $L_{D f_0}<L_{D^*_0\pi}$} $ , $3.4$ \\\hline
$\cpipim$ & $2.5$ , $3.0$ & $4.0$ , $2.9$\\\hline
$\cpid$ & $2.0$ , $0.5$ & $3.2$ , $4.0$\\
\hline\hline
\end{tabular}
\end{table}

It is interesting to examine the dependence of the $R$ value
on the decay mechanism.  The expression for
$R$ can be written as $s_1\times\br(B^-\to D_2^{*0} \pi^-, D_2^{*0} \to D^{*+}\pi^-)/$ 
$(s_1 \times \br(B^-\to D_1^0 \pi^-, D_1^0 \to D^{*+}\pi^-) +
s_2 \times \br(B^-\to D_1^0 \pi^-, D_1^0 \to D^0\pipi))$
where $s_i$ is a scale factor that recovers the full
width from the single decay channel.
(It includes branching fractions of other possible
decays of both $D^{**}$ and a meson from its decay products.)
Following the procedure used in Ref.~\cite{kuzmin} and fixing $s_1$ at $3/2$,
we can calculate $s_2$ factors for different models:
$s_2(D^{**0} \to D^{*+}_0\pi^-)=9/4$
(disregarding possible interference effects in
$D_1^0\to D^+\pi^-\pi^0$ decays),
$s_2(D^{**0} \to D^0 \rho)=3$, $s_2(D^{**0} \to D^0 f_0)=3/2$. 
Using the branching fractions measured in Ref.~\cite{kuzmin} and here,
the central value for $R$ depends on the
decay model: $0.50$ for $D\rho$, $0.60$ for $D f_0$ and
$0.54$  for $D^*_0\pi$.

The following sources of systematic errors are considered:
tracking efficiency (8\% overall, integrated over particle momenta), 
kaon identification efficiency (2\% overall),
$\pi^0$ reconstruction efficiency (8\%), $D$ branching fraction
uncertainties (2\%-7\%), MC statistics (2\%),
model uncertainty in MC efficiency (10\%),
uncertainty caused by variation of cuts (5\%),
background shape uncertainty (10\%).
The uncertainty in the tracking efficiency is estimated using 
partially reconstructed $D^{*+}\to D^0[K_S^0\pi^+\pi^-]\pi^+$
decays. The kaon identification uncertainty is determined from 
$D^{*+}\to D^0[K^-\pi^+]\pi^+$ decays. The $\pi^0$ reconstruction 
uncertainty is obtained using $D^0$ decays to $K^-\pi^+$ and 
$K^-\pi^+\pi^0$. 
To determine the systematic uncertainty in the signal yield
extraction, we use different parameterizations for the background
events.
The overall systematic uncertainty is 19\%
for $B\to D \pi\pi\pi$ and 21\% for $B \to D^*\pi\pi\pi$.
We assume equal production rates for $B^+B^-$ and $B^0\bar B^0$
pairs and do not include the corresponding uncertainty in the
total systematic error.

The $B^- \to D^0 \pi^+ \pi^- \pi^-$ final state also includes the
$D^{*+}\pi^-\pi^-$ intermediate state with $D^{*+}\to D^0\pi^+$.
We reverse the $D^*$ veto requirement to select $D^{*+}\pi^+\pi^-$ events and measure the branching ratio  $\br (B^- \to D^{*+} \pi^- \pi^-) = (1.27\pm0.07) \times 10^{-4}$
(based on a sample of $85 \times 10^6 B\bar{B}$ events),
that agrees well with the value of $\br (B^- \to D^{*+} \pi^- \pi^-) = (1.25\pm0.07) \times 10^{-4}$ measured earlier~ 
\cite{kuzmin}.

In summary, we report the first observation of $\done \to D\pipi$
decays  (with the dominant $D_1\to D^*\pi$ contribution excluded).
The measured branching ratios with the corresponding 
statistical significances and systematic uncertainties are
presented in Table~\ref{fitresults}. We find the upper limit for
the possible $D_2^*$ contribution to these results:
${\cal B}(B\to D_2^{*}\pi^-)\times{\cal B}(D_2^{*}\to D\pipi)
<
0.55 {\cal B}(B\to D_1\pi^-)\times{\cal B}(D_1\to D\pipi)$. No statistically significant signal has been observed for
the $D^{**} \to D^*\pipi$ decays. The corresponding 90\% CL upper
limits are listed in Table~\ref{fitresults}. 
Analysis of the
$D_1\to D\pipi$ dynamics shows that the decay model
$D_1 \to D_0^{*}\pi$ gives the best description of the data.
The 
$R={\cal B}(B^-\to D^{*0}_2\pi^-)/{\cal B}(B^-\to D^0_1\pi^-)$ 
value calculated assuming $D_1 \to D_0^{*}\pi$ dominates 
is $0.54\pm 0.18$; this is $\sim 2\sigma$ lower
than the previously published one.

   We thank the KEKB group for the excellent
   operation of the accelerator, the KEK Cryogenics
   group for the efficient operation of the solenoid,
   and the KEK computer group and the NII for valuable computing and
   Super-SINET network support.  We acknowledge support from
   MEXT and JSPS (Japan); ARC and DEST (Australia); NSFC (contract
   No.~10175071, China); DST (India); the BK21 program of MOEHRD and the
   CHEP SRC program of KOSEF (Korea); KBN (contract No.~2P03B 01324,
   Poland); MIST (Russia); MESS (Slovenia); NSC and MOE (Taiwan); and DOE
   (USA).

\begin{thebibliography}{99}
\bibitem{footnote1}{As most of the systematic uncertainties cancel in
the ratio of the branching fractions, the systematic uncertainty
is negligible.}
\bibitem{footnote2}{
We use the following expressions for matrix elements for
the $D_1\to D\pi\pi$ decay:
$
F^{\nu\eta}(B^0,\pi_B^+)
F_{\mu\nu}(\bar{D}_0^{*},\pi_{D_1^-}^-)
\cdot$ 
$P(D_1^-)_{\eta} P(D_1^-)^{\mu}
$,
$
F^{\phi\eta}(B^0,\pi_B^+)
F_{\mu\phi}(f^0, D^-)
\cdot$ 
$P(D_1^-)_{\eta} P(D_1^-)^{\mu}
$,
$
F^{\nu\eta}(B^0,\pi_B^+)
F_{\nu \mu}(\rho^0,\pi_\rho^+ - \pi_\rho^-)
\cdot$
$
P(D_1^-)_{\eta}P(D_1^-)^{\mu}
$,
where the following notation is used:
$F^{\mu\nu}(A,B) = P_A^\mu P_B^\nu - P_A^\nu P_B^\mu$,
$P(B-C)=P(B)-P(C)$
and $P$ stands for a 4-momentum.}
\bibitem{footnote3}{
In case of the $D f_0$ model 
the key $\cpid$ distribution is
practically independent of the $f_0$ 
mass and width.}
\end{thebibliography}
\end{document}